\font\bigg=cmbx10 at 17.3 truept    \font\bgg=cmbx10 at 12 truept
\font\twelverm=cmr10 scaled 1200    \font\twelvei=cmmi10 scaled 1200
\font\twelvesy=cmsy10 scaled 1200   \font\twelveex=cmex10 scaled 1200
\font\twelvebf=cmbx10 scaled 1200   \font\twelvesl=cmsl10 scaled 1200
\font\twelvett=cmtt10 scaled 1200   \font\twelveit=cmti10 scaled 1200
\def\twelvepoint{\normalbaselineskip=12.4pt
  \abovedisplayskip 12.4pt plus 3pt minus 9pt
  \belowdisplayskip 12.4pt plus 3pt minus 9pt
  \abovedisplayshortskip 0pt plus 3pt
  \belowdisplayshortskip 7.2pt plus 3pt minus 4pt
  \smallskipamount=3.6pt plus1.2pt minus1.2pt
  \medskipamount=7.2pt plus2.4pt minus2.4pt
  \bigskipamount=14.4pt plus4.8pt minus4.8pt
  \def\rm{\fam0\twelverm}          \def\it{\fam\itfam\twelveit}
  \def\sl{\fam\slfam\twelvesl}     \def\bf{\fam\bffam\twelvebf}
  \def\mit{\fam 1}                 \def\cal{\fam 2}
  \def\tt{\twelvett}
  \textfont0=\twelverm   \scriptfont0=\tenrm   \scriptscriptfont0=\sevenrm
  \textfont1=\twelvei    \scriptfont1=\teni    \scriptscriptfont1=\seveni
  \textfont2=\twelvesy   \scriptfont2=\tensy   \scriptscriptfont2=\sevensy
  \textfont3=\twelveex   \scriptfont3=\twelveex  \scriptscriptfont3=\twelveex
  \textfont\itfam=\twelveit
  \textfont\slfam=\twelvesl
  \textfont\bffam=\twelvebf \scriptfont\bffam=\tenbf
  \scriptscriptfont\bffam=\sevenbf
  \normalbaselines\rm}
\def\doublespace{\baselineskip=\normalbaselineskip \multiply\baselineskip by 2}

\newcount\equationnumber
\advance\equationnumber by1
\def\ifundefined#1{\expandafter\ifx\csname#1\endcsname\relax}
\def\docref#1{\ifundefined{#1} {\bf ?.?}\message{#1 not yet defined,}
\else \csname#1\endcsname \fi}
\def\autoeqnum{\def\eqlabel##1{\edef##1{\the\equationnumber}}}
\def\no{\eqno(\the\equationnumber){\global\advance\equationnumber by1}}
\newcount\citationnumber
\advance\citationnumber by1
\def\ifundefined#1{\expandafter\ifx\csname#1\endcsname\relax}
\def\cite#1{\ifundefined{#1} {\bf ?.?}\message{#1 not yet defined,}
\else \csname#1\endcsname \fi}
\def\autocite{\def\citelabel##1{\edef##1{\the\citationnumber}\global\advance\citationnumber by1}}
\def\preprintno#1{\rightline{\rm #1}}

\overfullrule = 0 pt

\hsize=6.5truein
\hoffset=.1truein
\vsize=8.9truein
\voffset=.05truein
\parskip=\medskipamount
\twelvepoint           
\doublespace
\autocite
\autoeqnum 


\def\cl{\centerline}

\vskip -48 truept
\preprintno{ICN-UNAM-96-09, adap-org/9610001}
{ \hfill October 8, 1996}\break

\cl{\bigg {SYMMETRY BREAKING AND ADAPTATION:}}
\cl{\bigg {THE GENETIC CODE OF RETROVIRAL ENV PROTEINS}}
\vskip 0.2truein
\cl{\bgg S. Vera Noguez}
\cl{\it Facultad de Ingenier\'\i a, UNAM,}
\cl{\it Circuito Exterior, C.U.,}
\cl{\it M\'exico, D.F. 04510.}
\cl{\bgg  H. Waelbroeck}
\cl{\it Instituto de Ciencias Nucleares, UNAM,}
\cl{\it Circuito Exterior, A.Postal 70-543,}
\cl{\it M\'exico, D.F. 04510.}
\cl{\it email: hwael@roxanne.nuclecu.unam.mx}

\noindent{\bf Abstract:}\ \ Although several synonymous codons can 
encode the same aminoacid, this symmetry is generally broken in 
natural genetic systems. In this article, we show that the symmetry 
breaking can result from selective pressures due to the violation of 
the synonym symmetry by mutation and recombination. We conjecture that 
this enhances the probability to produce mutants that are well-adapted 
to the current environment. Evidence is found in the codon frequencies 
of the HIV {\it env} protein: the codons most likely to mutate and 
lead to new viruses resistant to the current immunological attack, are 
found with a greater frequency than their less mutable synonyms. 

\vfill\eject

\line{\bgg 1. Introduction \hfil}

Darwin's proposal that evolution proceeds by random mutation and 
natural selection has been the keystone of evolution theory since 
the XIX'th century (Darwin 1859, Simpson 1964). Yet objections 
linger on: Large mutations require coordinated changes of several 
phenotypic traits, which seems unlikely to occur at random. 
Also the efficiency with which species adapt to changes in the 
environment has led some to suggest that there should be a mechanism for 
environmental feedback which favors useful mutations over random ones 
(Steele 1979). 

Proposals for a direct environmental feedback that would
pre-determine the mutations have been largely discarded: 
The so-called ``central dogma'' (Lewin 1995) states that information
from the environment cannot be tranfered to DNA. Actually this 
``dogma'' is not quite true, although the conclusion that there is no 
direct environment feedback probably {\it is}. 
Viruses can incorporate their own coding in the germ line cells, as inherited 
endogenous proviruses. The enzyme methylase can induce a mutation hotspot, 
which allows for indirect information transfer through the location of the 
hotspot, etc. Yet it seems difficult for information from the environment to be 
usefully transfered through such mechanisms, and this is why the central 
dogma is so well accepted. 

In this article, we will show that the environment
can organize the search of new genetic solutions {\it within the context
of random mutations of the chromosome}. The essential idea is that random
mutations of the {\it genotype} produce non-random mutations of the 
{\it phenotype}. This results from the existence of synonyms, together with
the claim that the gene pool breaks the synonym-symmetry in a way which 
incorporates information about the environment. 

 How this comes about is illustrated in the following simple example [Fig. 1].
Suppose that one has four possible genotypes, $A \to D$, and that
each genotype can mutate to the two adjacent genotypes when the letters
$A$, $B$, $C$, $D$ are placed on a circle. For example, $A$ can mutate
to $B$ or $D$ but not to $C$. $A$ and $D$ both encode the 
phenotype $a$, $B$ encodes $b$ and $C$ encodes $c$. In a random population, 
$p(A) = \cdots = p(D) = {1 \over 4}$ and the phenotype distribution 
is $p(a) = {1 \over 2}$, $p(b) = p(c) = {1 \over 4}$. The distribution of 
mutant phenotypes in that case is the same as for the population prior to
mutation. But if the gene pool is organized so 
that phenotype $a$ is always encoded as $A$, then with the same distribution 
of phenotypes one gets the mutant distribution $p(a) = {1 \over 2}$, 
$p(b) = {1 \over 3}$ and $p(c) = {1 \over 6}$. Vice-versa, if the synonym 
$D$ is chosen instead of $A$ then the mutations to the phenotype $c$ are
more probable than to $b$. Information in the genotype probability 
distribution is expressed through a non-random distribution of mutant 
phenotypes. 

 The next step is to explain how selective forces can induce the symmetry
breaking of the gene pool, thereby incorporating information about the
environment into the genotype distribution. 

 Symmetry breaking would occur spontaneously in a finite breeding pool, 
by the theory of branching processes (Taib 1994, Garc\'\i a-Pelayo 1994)
(this observation is the backbone of the Neutral Theory of molecular 
evolution (Kimura 1983). However, we will show 
that there is also an {\it induced symmetry breaking} from the violation 
of the synonym symmetry by the genetic operators, such as mutations and 
recombination. 

 If one considers the growth of an allele not from one generation to the next
but over many generations, selection forces will take into account not only
the selective advantage of this allele but its ability to produce 
well-adapted offspring, which can themselves produce well-adapted offspring,
etc. Since mutation and recombination act differently on synonymous alleles, 
the synonyms will differ in their descendence, both in the passive 
sense of genes surviving mutations, and in the active 
sense, to generate new genetic solutions. This implies that the time-averaged 
{\it effective fitness} function, defined as the growth rate of an allele over 
many generations, does not respect the synonym symmetry.

 Thus, that the time-averaged
effective fitness function provides a selective pressure which enhances the 
production of potentially successful mutants by selecting, among the synonyms, 
those that have a higher probability to generate well-adapted offspring. 

 This is a highly non-trivial proposal, in that it implies an 
environmental feedback in the genetic search: if the symmetry breaking is
due in part to selective pressures, then the section of the symmetry 
group naturally incorporates information about the current environment, so that
mutant phenotypes are produced not at random, but to some extent  
tuned to the current environment. Although this suggestion brings back the
ghost of Lamarckism (Simpson 1964), we stress 
that what we are saying is in no way in
contradiction with the central dogma. Information from the environment is
incorporated indirectly through the symmetry breaking of the gene pool, not 
at the level of a single individual.

 The simplest manifestation of synonym symmetry in the genetic code is
the codon redundancy. Since $4^3 = 64$ possible codons encode 20 aminoacids
and a $STOP$ sequence, most aminoacids can be represented by several  
codons. Synonymous codons differ in terms of their products following 
a mutation of a single nucleotide. 

 It is easy to see how different synonyms can have different mutabilities.
Consider for example the synonymous words {\it dead}
and {\it defunct}, and assume that a mutation changes a single letter. 
The word {\it dead} can mutate to {\it deed}, {\it bead},
{\it lead}, {\it deaf}, {\it dean}, {\it dear}, {\it read} or {\it deal}, 
but it is difficult to generate a meaningful word by mutating 
the word {\it defunct}. As we will see below, the situation 
with codons is similar. The time-averaged effective fitness
will give a selective edge to codons with the ability to mutate to another 
useful form. Since what is ``useful'' is generally environment-dependent, 
this implies that the symmetry-breaking process incorporates information 
about the environment into the gene pool. 

 In this paper, we will consider the code of retroviral enveloping 
proteins, which are a key component in the detection of a virus by 
the immune system. We will show that the codons which have a higher probability 
to mutate to a different allowed aminoacid are prefered, thereby enhancing 
the ability to escape detection by the immune system. This demonstrates that 
phenotype mutations are organized, and furthermore that this 
organization is guided by environmental factors. They are {\it organized} 
because certain mutant aminoacids are produced more frequently than with 
a random choice of codons, and {\it guided by environmental factors} 
because these more frequent aminoacids are precisely the ones that are 
allowed (and thus presumably lead to a successful infection). 

 In order for this mechanism to usefully guide the search of new genetic
solutions in a more complex organism, one must generalize the concept of ``synonym''
beyond the single-codon degeneracy of the genetic code. The central concept 
which allows selection among synonyms to take place is that of indirect 
encoding. The chromosome does not encode directly the size and shape 
of various parts of an organism, but an {\it interpreter}, embodied by 
the biochemical processes in living cells, translates the genotype 
into a phenotype. The decoder allows for several types of synonyms. The
most trivial example is that of various codons encoding the same amino-acid, 
but there are more subtle synonyms, related to the machinery of gene expression, 
for which symmetry breaking can be related to the emergence of an {\it algorithmic
language}.

 If one views the chromosome as an algorithm, the interpreter is the ``computer'' 
which executes the algorithm and the phenotype is the solution. In 
this sense, the breaking of synonym symmetry is related to the selection of
a language, where ``words'' or ``gramma-tical rules'' are selected if they facilitate 
the search for successful mutants. This will be the case if they are related to
an approximate decomposition of the optimization problem into smaller subproblems.
This in turn requires that the genetic interpreter be sufficiently flexible to realize the
required decomposition, and that the fitness landscape be sufficiently correlated 
to allow for a decomposition of the optimization problem into smaller subproblems. 
An example would be Kauffman's $Nk$ landscapes for $k << N$, together with 
his model of cellular gene regulation (Kauffman 1993).

The importance of viewing the genotype as an algorithm for a solution 
rather than the solution itself has been discussed previously in
the context of genetic algorithms (Asselmeyer et al. 1995, Adami 1994), as 
has the idea that intelligence is an emerging collective property 
(e.g., Rauch et al. 1995). Our claim is that the 
existence of synonyms and a mechanism of symmetry breaking are the 
key to understanding how such ideas are realized in practice. 

Since the algorithmic language is necessarily evolved in the causal past, 
it will facilitate the search of {\it future} solutions only if the decomposition
of the optimization problem is stable. We will refer to
this condition as {\it structural decomposition stability}: 
the evolution of the landscape must preserve the structural decomposition 
of the adaptation problem, so that the language which was successful in 
the past continues to be successful in the immediate future. 

One might conjecture that extinctions are related to a violation of  
structural decomposition stability. For instance, the algorithmic language 
guiding the search of new dinosaur species would presumably have been 
incapable of producing viable solutions in the environment which is 
assumed to have provoked their demise. 

In the following sections, we will focus on the simplest possible example
of symmetry breaking in the genetic code, in order to demonstrate both
analytically and experimentally the concept of {\it induced symmetry breaking}.
We should stress that our results by no means exclude the possibility of 
{\it spontaneous} symmetry breaking, which is postulated in the Neutral 
Theory and strongly backed by empirical evidence. Likewise, our proposal 
and Cocho and Mart\'\i nez-Mekler's views on symmetry breaking by the 
genetic transcription machinery (Cocho et al. 1992, 1994, 1995) are 
not mutually exclusive. 

We will consider the synonymous codons at each amino-acid in the {\it env} 
protein from an HIV database (Myers 1992). The synonymous 
codons differ in their products following a point mutation. Some of these 
products are ``allowed'', in the sense that they give rise to a functional 
virus, others may be neutral (same amino-acid), or forbidden. We will 
show that codons with a higher probability to mutate to a different 
allowed amino-acid are favored over their less ``mutable'' synonyms. 
This enhances the probability that the virus generate mutants capable 
of escaping detection by the immune system.

\
 
\line{\bgg 2. Symmetry breaking through codon mutability \hfil}

 The data on {\it env} proteins gives sequences which can be aligned 
in the usual way to identify a total of 978 codon positions, some of which
may be unoccupied in a particular strain of the virus. Let us focus on 
one position in particular, and make a list of the different aminoacids
that are found at that position: these will be called ``allowed 
aminoacids'', by definition. We will be interested in the effect of 
point mutations in the transcription of the {\it env} protein in the 
HIV-1 viruses from our database. 

 Some point mutations lead to a synonymous codon which represents 
the same aminoacid; we will refer to such point mutations as {\it neutral}. 
When the mutation leads to a {\it different} amino-acid, we will say that
the non-neutral mutation is ``allowed'' if the aminoacid obtained is
an ``allowed aminoacid'', i.e. if it can be found in another sequence 
at the same position. 

 For a given codon there are 9 possible point mutations. We will call
the ``mutability'' of a codon the number of allowed mutations. When 
only non-neutral allowed mutations are counted one gets the ``proper 
mutability''. Following our reasoning in the introduction, 
one expects that the codons with the highest mutability will be selected, 
since they have a higher probability to sucessfully infect a 
cell. The codon with a high {\it proper mutability} has an extra selective 
advantage since its offspring may be able to fool the immune system, 
if the mutation happens to be picked out by macrophages and displayed 
as part of an epitope. Thus, statistically speaking, codons with a 
higher mutability are selected for resistance to mutations and those 
with a high proper mutability are selected for adaptation, in this 
case to the immune system. 

 For example, at position number 186 one finds
the aminoacids leucine, which can be encoded by (TTA,TTG,TCT,TCC,TCA,TCG), 
glutamic acid (AGG,AGA), triptophan (GTG) and {\it STOP} (ATA,ATG,GTA), i.e.
12 possible codons for four allowed aminoacids. The mutability and
proper mutability of each codon are represented in [Table 1]. Note 
that synonyms differ in terms of their mutabilities and proper mutabilities: 
For example, TTA can mutate in three different ways to the 
{\it STOP} sequence, but its synonym TCT allows no 
non-neutral mutations.

 Let us define the {\it frequency} of a codon as the number of times 
this codon is found at that position, as one scans through the sequences 
in the database. If selection is behind the symmetry breaking phenomenon,
as we have speculated, the more mutable codons should have a higher 
frequency. 

 To represent the relation between the frequency with which codons are 
observed and their mutabilities, we computed the linear correlation 
coefficient between frequency and mutability at every position where 
the variance is non-zero, and likewise for frequency and proper mutability. 
The correlation coefficients are represented in the form of a histogram
in both cases. In the case of mutability [Figure 2a] there seems to
be a preference for positive correlation coefficients: there are 114 
positive correlation coefficients against 42 negative ones. 
The next histogram however is even more dramatic:  
The weight of the histogram [Figure 2b] is clearly shifted towards
positive correlations, a result which would be difficult to 
explain without recognizing that codons with a higher {\it proper 
mutability} are prefered. 

 Considering the importance of proper mutability to help the virus adapt 
to the attacks of the immune system, these results are evidence that 
natural genetic systems can use symmetry breaking to organize the 
distribution of mutants. As we speculated in the Introduction, 
the genetic search through random mutations of the genotype produces an 
intelligent search at the level of mutant {\it phenotypes}, as the 
information in the probability distribution of codons is revealed through 
an better-than-random probability distribution of mutant phenotypes.

 In some positions one finds a negative correlation with mutability but 
a positive correlation with proper mutability. This may imply that non-neutral
mutations at such a site are important to escape the immune system, i.e.
that the site is frequently picked as part of an epitope. The adaptability 
criterion would then overrule the slight selective advantage of resistance 
to mutations. Vice-versa, those sites with a positive correlation to mutability
but negative correlation to proper mutability may simply never be picked
as part of epitopes. Negative or null correlations can also indicate 
violations of the assumptions underlying our definitions of ``allowed
aminoacids'' and ``allowed mutation'', namely that codons act 
independently of each other and that our dataset is complete, or 
simply be the result of statistical fluctuations, e.g. neutral 
drift.

\

\line{\bgg 3. A simple model for symmetry breaking\hfil}

 In this section we will provide a simple model which allows one to
calculate the value of the correlation coefficients represented 
in [Figure 2]. The model relies on highly simplistic assumptions, so
our purpose here is to obtain an order of magnitude for the expected 
correlation coefficient, not a precise result. 

 We will assume that any one of the three bases which constitute the codon 
is replaced at random. Since there are four possible bases, the 
mutation probability is $p = 1/3$ to each of the other bases. Considering
the three bases which constitute a codon, the probability of each point
mutation will then be $1/9$. We will further assume that one can analyze 
each codon position independently of the others. 
The probability that a virus in the population successfully infect a 
cell and thereby perpetuates itself will be taken to be {\it one}
if the transcribed codon is allowed and {\it zero} if it is not. So
the ``fitness'' is a binary number. Clearly these assumptions are too
simplistic for a serious model of a virus; however, they can be expected to
provide a rough upper bound for the expected correlation coefficient: 
Indeed, non-local effects related to the ``cooperation'' of different codons, 
or restrictions on the genotype mutations, all dilute the effect
which we are trying to observe by reducing the amount of information 
available to organize the phenotype mutations and by bringing in other
competing motives for symmetry breaking. 

 The codons which represent allowed aminoacids will be labeled by 
a latin index $i = 1, 2, \cdots, N$, where $N$ is the total
number of allowed codons at that site. Their mutability 
will be denoted by $m_i$. A point mutation is assumed to occur at the 
given position with probability $p$. 

The equation which determines the evolution of the frequency of a codon 
in the population given these assumptions is

$$n_i(t+1) = (1 - p) n_i(t) + {p \over 9} \sum_{j=1}^N M_{ij} n_j(t) ,$$
where $n_i(t)$ is the number of times that an individual in the population has
the codon $i$ at that site, and $M_{ij} = 1$ if $i$ can mutate to $j$ with a point 
mutation and zero otherwise. The {\it mutability} of codon $i$ is given by 
$m_i = \sum_j M_{ij}$.

 In the limit $p \to 0$ and taking an infinite population $P \to \infty$, 
$$P_i(t) = {n_i(t) \over {\sum_{j=1}^P n_j(t)}}$$
is well-approximated by a real variable and the equation above becomes 
equivalent to the differential equation

$${d P_i \over dt} = p \sum_{j=1}^N N_{ij} P_j(t),$$
where $N_{ij} = M_{ij} - \delta_{ij}$. This is set of linear differential 
equations. The population has an asymptotic conformal fixed point 
as $t \to \infty$ given by the eigenvector of the matrix ${\bf N}$ with 
the largest eigenvalue, or ``Liapunov exponent''. 

 Given this eigenvector one can compute the correlation between the 
mutability and the frequency of a codon in the population, defined by

$$C = \lim_{t \to \infty} \ {1 \over N} {{\sum_{i = 1}^N \ 
(m_i - {\buildrel \_ \over m} ) ( P_i(t) - 
{\buildrel \_ \over P} )} \over 
\sigma_m \sigma_P},$$ 
where $\sigma_m$ and $\sigma_P$ are the root mean squares of mutability 
and codon frequency, respectively, and ${\buildrel \_ \over m}$, 
${\buildrel \_ \over P}$ their mean values. 

 Several correlation coefficients were computed using this model and gave
values which fluctuate about the mean of the observed correlation 
coefficients. For example, at position 186 one finds the aminoacids leucine, 
glutamic acid, triptophan and {\it STOP}, i.e. 12 possible codons with
the mutability matrix given in [Table 1]. The eigenvector which 
corresponds to the least negative eigenvalue, $\Lambda_0 = -5.19$, is 
given. This eigenvector gives the asymptotic value for the frequency 
of the various codons in the population according to our simple model. 
The relation between mutability and codon frequency is manifest and 
gives a correlation coefficient of 0.164. This does not compare well 
to the observed correlation at this position, which was anomalously high 
at 0.74, but is the same order of magnitude as the mean observed 
correlation coefficient.

 A similar model can be constructed to examine the correlation with 
{\it proper} mutability. The only required modification is
in the definition of the fitness function: the probability of successful
infection, or ``fitness'', is assigned a larger value if the allowed
mutation is {\it non-neutral}. 

 Note that the arguments given above are reversible: if sections of 
genetic code are best kept {\it invariant} then the favored codons would be
the less mutable ones. This effect would be represented in a model 
which assigns fitness zero to non-neutral mutants.  

\

\line{\bgg 4. Statistical evidence for a coding language\hfil}

 The results of section 2 suggest that there is a {\it language} in the
genetic code which enhances the production of non-neutral mutations of 
the {\it env} protein.  

 Various statistical tests are available to check whether a sequence 
can be based on a coding language; all have been applied recently to 
search for evidence of a language in both coding and non-coding DNA. We 
shall mention only a few of them. Zipf analysis with fixed word-length 
yields the exponent $\zeta$ (Mantegna 1995), 
the Shannon entropy (Crisanti et al. 1993) is usually 
represented by giving percentage of redundancy defined by
$$R(l) = 100 \times \left( 1 - {H_1(l) \over ln(4^l)} \right).$$
The ``digital walks'' (Peng et al. 1992, Buldyrev et al. 1993) yield 
the correlation exponent $\alpha$ which describes the scaling of 
the variance (Peng et al. 1992, Li and Kaneko 1992), as well as some 
other less-used exponents (Voss 1992). 
In [Table 2] we give a few of the results of these analyses, taken 
from the references in this paragraph.

 There is some controversy in the literature regarding these methods
(Israeloff et al. 1996, Bonhoeffer et al. 1996, Voss 1996, Mantegna et al. 
1996). The results of Zipf analysis and entropy measurements can be 
reproduced with stochastic sequences; 
also the binary walks are similar to those from generalized
Levy walks (Buldyrev et al. 1993). The bottom line seems to be
that any {\it statistical} measure of a language can be reproduced  
with a stochastic sequence tailored to have the appropriate characteristics.
This should come as no surprise: statistical methods can only test 
statistical properties, and one can always construct a stochastic 
sequence with any desired statistical properties. A {\it language} 
presumably involves something more than statistics, namely it 
requires the assumption of an underlying intelligence. The only way to 
prove the existence of a language, then, is to unravel its meaning!

 Statistics may provide {\it evidence} for a language, but for a {\it proof} 
one must demonstrate the existence of a meaning. This is essentially
what we have attempted to do in this article. 

 In any case, to supplement our results we have carried out these 
standard statistical tests. The results are summarized in [Table 2]
and in [Figs. 3, 4]; they are consistent with the possibility of a 
coding language for viral RNA.

\

\line{\bgg 5. Conclusions \hfil}

 We have proposed a mechanism whereby the environment can organize the
genetic search by condensing the gene pool into an ``intelligent'' broken
symmetry phase. This would optimize the probability that mutant offspring
are well adapted to the current environment. 

 To search for evidence that such a mechanism is at work, we considered the
enveloping protein of the retrovirus HIV, and found that codons with a 
high mutability are more frequent than their less mutable synonyms. Results 
from an analysis of genetic data show positive correlation coefficients,
especially between proper mutability and codon frequency. The observed
correlation coefficients are of the same order of magnitude as those 
derived from the theoretical model, and the average value $C \approx 0.25$ 
is consistent with expectations.

 In the language of optimization theory and artificial life, mutability
reduces brittleness, while proper mutability organizes the generation of
mutants as an ``intelligent'' search. Our results suggest that for viral
populations the organization of mutant production to adapt to the
immune system is more important than reducing brittleness.

 The results also confirm the importance of non-neutral mutations 
in the dynamics of the AIDS virus. This appears to support a conjecture by 
Nowak (Nowak 1992) concerning the latency period of 
the virus prior to the manifestation of AIDS: Nowak has argued 
that the number of mutant forms of the virus increases until the
immune system is overwhelmed by the uneven battle between the specific immune
response, which must target every one of the mutant viruses individually,
and the non-specific attack of the virus on the CD4 cells. Our arguments
concerning the self-organization of the gene pool also 
suggest what may be another part of the explanation for the long latency
period: If the gene pool can organize to enhance the mutation rate of the 
epitopes which provoke the strongest immune response specific to the
particular infected individual, this would effectively 
disable the strongest sector of the individual's immune system. Preliminary 
numerical simulations with the help of genetic algorithms indicate that the
timescale for adaptation of the gene pool to the specificities of an 
individual's immune system is of the order of several thousand generations 
of viruses. The mutation rate of HIV is consistent with a rate of roughly
one thousand generations per calender year. Therefore, observed latency 
periods give the right amount of time to evolve the symmetry-broken phase
of the gene pool, to adapt to the individual's MHC and epitopal response
specificities. 

 The relation between symmetry breaking of the gene pool and adaptability
of the virus may also have implications in disease control and therapy. For 
example, the identification of a segment of DNA dominated by codons with a 
low proper mutability would naturally signal a possible conserved region. 
This can be useful both with regard to detection tasks, through the 
polymerase chain reaction (PCR) (Lewin 1995), or in vaccine design, 
if a conserved part of the envelope of a dangerous virus can be incorporated 
in that of a controlled one (e.g. smallpox), to provoke a persistent induced 
immune response. From this perspective, it would be 
interesting to consider the reversed situation, where conserved regions 
would be identifiable by their preference for non-mutable codons over 
mutable ones. 

 Our results show that induced symmetry breaking does in fact play a role in
molecular evolution, but falls short of actually demonstrating environmental
feedback and the emergence of a more sophisticated form of intelligence 
in the genetic search, that would be non-local along the chromosome. This
more ambitious proposal is difficult to check empirically, due to our 
lack of a detailed understanding of the genetic interpreter. For this
reason we are currently developing genetic algorithms based on Kauffman's 
model of the genetic interpreter (Kauffman 1993) and other indirect 
encoding models, with which we hope to demonstrate the possibility 
that an algorithmic language emerge from the symmetry breaking.  

\

\noindent {\bf Acknowledgements} We would like to express our gratitude
to Ricardo Mancilla for his valuable comments on the manuscript, and 
Germinal Cocho and Gustavo Mart\'\i nez-Meckler for conversations and 
access to the data. The early developement of this 
work was strongly influenced by conversations with Ricardo Garc\'\i a-Pelayo.

\vfill\eject

\

\centerline{\bf REFERENCES}

\

\item{} Adami, C., 1994: {\it Learning and Complexity in Genetic 
Auto-Adaptive Systems}, Preprint adap-org/9401002 from nlin-sys@xyz.lanl.gov.

\item{} Asselmeyer, T., Ebeling, W. and Ros\'e, H., 1995: {\it 
Smoothing Representation of Fitness Landscapes -- the Genotype-Phenotype
Map of Evolution}, Preprint adap-org/9508002 from nlin-sys@xyz.lanl.gov,
submitted to Bio. Sys.

\item{} Bonhoeffer, S. et al., 1996:
Phys. Rev. Lett. {\it 76} No. 1 , 1977 (comment).

\item{} Buldyrev, S. V. et al., 1993: Phys. Rev. {\bf E  47}
, 4514.

\item{} Cocho, G. and Mart\'\i nez-Mekler, G., 1995: {\it Modeling
the AIDS Virus Genetic Evolution: From Nonlinear Dynamics to Molecular
Machines}, in: ``CAM 94 Physics Meeting, Canc\'un, M\'exico 1995), AIP 
Conference Proceedings 342, A. Zepeda, editor (AIP Press, Woodbury, 
New York, 1995) pp. 679-694.

\item{} Cocho, G., Gelover-Santiago, A. Mart\'\i nez-Mekler, G. and 
Rodin, A., 1994: International Journal of Modern Physics C , 321-324.

\item {} Cocho, G. et al., 1992: 
Rev. Mex. Fis 28, suplemento 1 , 127.

\item{} Crisanti, A., Falconi, M. and Vulpiani, A., 1993:
J. Phys. {\bf A 26} , 3441. 

\item{} Darwin, C., 1859: {\it The Origin of Species by Means of Natural 
Selection}, John Murray, London.

\item{} Garc\'\i a-Pelayo, R., 1994: Phys. Rev. {\bf E 49} , 4903

\item{} Israeloff, N. E. et al., 1996:
Phys. Rev. Lett. {\it 76} No. 1 , 1976 (comment).

\item{}  Kauffman, S. A., 1993: {\it The Origins of Order}, 
Oxford University Press, Oxford .

\item{} Kimura, M. , 1983: {\it The Neutral Theory of Molecular
Evolution}, Cambridge University Press, Cambridge .

\item{} Mantegna, R. N. et al., 1995: Phys. Rev. {\bf E 52}
, 2939.

\item{} Mantegna, R. N.  et al., 1996:
Phys. Rev. Lett. {\it 76} No. 1 , 1979-1981 (reply).

\item{} Myers, G. L.  et al., 1992: {\it Human Retroviruses and
AIDS} (Los Alamos National Laboratory, Los Alamos, New Mexico) 
[Fall Edition, HIV database]

\item{} Lewin, B., 1995: {\it Genes V}, Oxford University Press, Oxford 

\item{} Li, W.  and Kaneko,K. , 1992:  Europhys. Lett. {\bf 17}
, 655.

\item{} Nowak, M. A. , 1992: J. Theor. Biol. {\bf 155}, 1

\item{} Peng,C. K.  et al., 1992: Nature {\bf 356} , 168.

\item{} Rauch,E. M. , Millonas, M. M.  and  
Chialvo, D. R., 1995: {\it Pattern Formation and Functionality in Swarm Models}, 
Preprint adap-org/9507003 from nlin-sys@xyz.lanl.gov, submitted to
Physics Letters A .

\item{} Simpson,G. G. , 1964: {\it This View of Life}, Harcourt, Brace 
\& World, New York .

\item{} Steele,E. J. , 1979: {\it Somatic Selection and Adaptive Evolution},
Williams Wallace, Toronto.

\item{} Taib,Z., 1991 : {\it Branching Processes and Neutral 
Evolution}, Springer Verlag, Berlin  .

\item{} Voss, R. F. , 1992: Phys. Rev. Lett. {\bf 68} , 3805.

\item{} Voss, R. F. , 1996:
Phys. Rev. Lett. {\it 76} No. 1 , 1978 (comment).

\vfill\eject

\
\centerline{\bf FIGURE CAPTIONS}

\

\item{} Figure 1. The genetic interpreter translates a genotype (left)
to a phenotype (right). When this is a surjection, the domain set of
a particular phenotype consists of ``synonyms''. Different synonyms
may differ in terms of their mutability; when that is the case, 
mutant {\it phenotypes} can be predetermined to some extent by choosing 
among the equivalent codes for each phenotype.   

\

\item{} Figure 2. The correlation coefficients between codon mutability 
and codon frequency (a), and between proper mutability and frequency (b), 
are represented. The vertical axis gives the 
number of correlation coefficients in each interval, out of the 978 
positions of the HIV {\it env} segment. Since most positions allow only
one possible aminoacid, the total number of correlation
coefficients computed is less than 978.

\

\item{} Table 1. The mutability matrix is given for the 12 possible
codons at position 186 of the HIV {\it env} segment, considering only 
single point mutations. The total mutability of each codon and the 
dominant eigenvector of the matrix $I - {1 \over 9} M$ are given in 
the last two columns. The dominant eigenvector represents the 
theoretical expected values for the asymptotic 
frequencies of each codon in the population as $t \to \infty$. The 
calculation is based on a single-codon model where the offspring 
survives if and only if the inherited codon encodes an allowed aminoacid.

\

\item{} Table 2. The correlation exponent $\alpha$, the Zipf analysis 
exponent $\zeta$ and the Shannon redundancy $R(4)$ are given for various
examples of genetic code. These figures are in general agreement with the
conjectured existence of a coding language.

\

\item{} Figure 3. The Zipf plots are given for the coding of HIV {\it env} 
proteins. Words of a given length are ranked from the most probable to
the least probable, for word lengths ranging from 3 to 6. The data 
is not quite sufficient to obtain an accurate evaluation
of $\zeta$, but a power law behavior is observed over $1 {1 \over 2}$ 
decades in all cases. 

\

\item{} Figure 4. The Shannon redundancy $R(l)$ is given for $l = 1, 2, 3, 4$,
for the coding of HIV {\it env} proteins. This gives a measure of the 
percent reduction of entropy as compared to a random sequence. 

\end